\documentclass[aps,prd,twocolumn]{revtex4}
\usepackage{amsmath,amssymb}

\newcommand{\eqa}{\begin{eqnarray}}
\newcommand{\neqa}{\end{eqnarray}}
\newcommand{\be}{\begin{equation}}
\newcommand{\ee}{\end{equation}}

\renewcommand{\texttt}{{}}
\usepackage{graphicx}

\begin{document}
\title{\large Fractal Dimension in 3d Spin-Foams} % Models} 
%\title{Spectral Dimension in 3d Quantum Gravity} %Spinfoam} 
%\title{Fractal Structure of 3d Quantum Gravity} 
%\title{Fractal Anatomy of 3d Quantum Space-Time}%SpinFoams}
%\title{Spectral Dimension from 3d Spinfoams} % Quantum Gravity}%Models}%Quantum Space-Time}
%\title{Fractal Dimension in 3d Quantum Gravity} 
%\title{Fractal Structure of 3d Quantum Gravity} 
%\title{Fractal Structure of Non Perturbative 3d Quantum Gravity} 
%\title{Fractal Properties of the Ponzano-Regge Model} %Structure of 3d Quantum Gravity} 
%\title{Fractal Properties of a 3d Spinf-Foam Model} %Quantum Gravity} 
\author{Francesco Caravelli$^\ast$ %\footnote{fcaravelli@perimeterinstitute.ca} 
\& Leonardo Modesto$^\dagger$%\footnote{lmodesto@perimeterinstitute.ca}
}
%\vspace{0.4cm}
%\affiliation{$^\ast$$^\dagger$Dipartimento di Fisica 'E. Fermi', Pisa, Italy} 
%\\\&\\ Perimeter Institute for Theoretical Physics, Waterloo, ON, Canada}
%\author{Leonardo Modesto\footnote{lmodesto@perimeterinstitute.ca}}
\affiliation{$^\ast$$^\dagger$Perimeter Institute for Theoretical Physics, 31 Caroline St., Waterloo, ON N2L 2Y5, Canada}
\affiliation{$^\ast$Universit$\grave{{\rm a}}$ di Pisa, Italy}

\date{\small\today}
%integrand 

\begin{abstract} \noindent
In this paper we perform the calculation of the spectral dimension of the space-time in 3d quantum gravity
using the dynamics of the Ponzano-Regge vertex (PR) and its quantum group generalization
(Turaev-Viro model (TV) \cite{turviro}). 
We realize this 
considering a very simple decomposition of the 3d space-time and 
introducing a boundary state which selects a classical geometry on the boundary.
We obtain that
the spectral dimension of the space-time runs from $\approx 2$ to $3$, across a $\approx 1.5$ phase, 
when the energy of a probe scalar field decreases from high $E \lesssim E_P$ to low energy. 
For the TV model the spectral dimension at hight energy increase with the value of the cosmological constant $\Lambda$. At low energy the presence of $\Lambda$ does not change the spectral dimension.
%We have calculated 
%also the spectral dimension of the space-time using results from spin-foam models and 
%obtained a $2$-dimensional effective manifold at high energy. Our result is consistent with two 
%other approaches to non perturbative quantum gravity: 
%{\em causal dynamical triangulation} 
%and {\em asymptotically safety quantum gravity}.

\end{abstract}
\maketitle
{\em Introduction.} In past years many approaches to quantum gravity studied the fractal properties of the quantum space-time. 
In particular in {\em causal dynamical triangulation} (CDT) \cite{CDT} 
and {\em asymptotically safe quantum gravity} (ASQG) \cite{ASQG}, a fractal analysis 
of the space-time gives a two dimensional effective manifold
at high energy. In both approaches the spectral
dimension is ${\mathcal D}_s=2$ at small scales and ${\mathcal D}_s=4$ at large scales.%%%%%%%
The previous ideas have been applied in the context of {\em non commutativity} 
to a quantum sphere and
$\kappa$-Minkowski \cite{Dario} and in Loop Quantum Gravity \cite{Modesto0}.
%. In particular, for the second group the author 
%found a space-time spectral dimension ${\mathcal D}_s=3$ at high energy. 
The spectral dimension has been studied also in the cosmology of a Lifshitz universe
\cite{calcagni} and in Causal Sets \cite{DR}.
Spectral analysis is a useful tool to understand the {\em effective form}
 of the space at small and large scales. 
We believe that the fractal analysis could be also a useful tool to predict the behaviour 
of the $2$-point and $n$-point functions at small scales \cite{propaga} and to attack the 
singularity problems of general relativity in a full theory of quantum gravity \cite{CBH}. 

In this paper we apply to the %spinfoam model %{\em loop quantum gravity}
 Ponzano-Regge (PR) model \cite{PR} and to the Turaev-Viro model (TV) \cite{turviro}, \cite{book1}, \cite{book2} the analysis introduced in \cite{Modesto0}.
%developed in the context of ASQG 
%by O. Lauscher and M. Reuter \cite{ASQG}.
We consider the appropriate spinfoam model and we use 
the very simple decomposition of the $3d$ space-time introduced by
Speziale in \cite{speziale}. The other ingredient is the general boundary formalism 
useful to define the boundary geometry \cite{gb}. 
All the space-time is approximated by a single tetrahedron 
and the boundary state is peaked on the boundary geometry of it.

The paper is organized as follows. In the first section we define the framework and 
we recall the definition of spectral dimension in quantum gravity.
The analysis in this section is general and not strongly related to the 
PR or TV models. The analysis is correct for any spin-foam model.
In the second section we 
calculate explicitly the spectral dimension for the PR and the TV theories using
the general boundary formalism to define the 3d quantum gravity path integral. %\\\ \\

{\em The Spectral Dimension}.
%\section{The spectral dimension}
%\renewcommand{\theequation}{\arabic{equation}}
%\setcounter{equation}{0}
%\label{Sec3}
%In this section we define the spectral dimension ${\cal D}_{\rm s}$ 
%of the quantum space-time.
The following definition of a fractal dimension is borrowed
from the theory of diffusion processes on fractals \cite{avra} and is easily 
adapted to the
quantum gravity context. %\cite{nino,ajl34}. In particular it has been used in
%the Monte Carlo studies mentioned in the Introduction.
Let us study the diffusion of a scalar test (probe) particle on a $d$-dimensional 
classical Euclidean manifold with a fixed smooth metric $g_{\mu\nu}(x)$.
The corresponding heat-kernel $K_g(x,x';T)$ giving the probability for the
particle to diffuse from $x'$ to $x$ during the fictitious diffusion time $T$
(this is just a fictitious time and the scalar field in general is just a tool to understand
the fractal properties of the space-time)
satisfies the heat equation
\begin{eqnarray}
\label{heateq}
\partial_T K_g(x,x';T)=\Delta_g K_g(x,x';T)
\end{eqnarray}
where $\Delta_g$ denotes the scalar Laplacian: 
$\Delta_g\phi\equiv g^{-1/2}\,\partial_\mu(g^{1/2}\,g^{\mu\nu}\,\partial_\nu
\phi)$. The heat-kernel is a matrix element of the operator 
$\exp(T\,\Delta_g)$, 
\begin{eqnarray}
K_g(x,x';T) =\langle x^{\prime} | \exp(T\,\Delta_g) | x \rangle.
\label{EK}
\end{eqnarray} 
In the random walk picture its trace per unit volume,
\begin{eqnarray}
\label{trace}
&& P_g(T)\equiv V^{-1}\int d^dx\,\sqrt{g(x)}\,K_g(x,x;T) \nonumber \\
&& \hspace{1cm} \equiv 
V^{-1}\,{\rm Tr}\,
\exp(T\,\Delta_g)\;,
\end{eqnarray}
has the interpretation of an average return probability. (Here $V\equiv\int
d^dx\,\sqrt{g}$ denotes the total volume.) It is well known that $P_g$
possesses an asymptotic expansion (for $T\rightarrow 0$) of the form
$P_g(T)=(4\pi T)^{-d/2}\sum_{n=0}^\infty A_n\,T^n$. For an infinite flat
space, for instance, it reads $P_g(T)=(4\pi T)^{-d/2}$ for all $T$. Thus
from the knowledge of the function $P_g$ one can recover the dimensionality of the target
manifold as the $T$-independent logarithmic derivative
\begin{eqnarray}
\label{dimform}
d=-2\frac{d\ln P_g(T)}{d\ln T}
\end{eqnarray}
This formula can also be used for curved spacetimes and spacetimes with finite volume
$V$ provided that $T$ is not taken too large. % \cite{ajl34}.

In quantum gravity % where we functionally integrate over all metrics 
it is natural to replace $P_g(T)$ by its expectation value %on the spin-network states
on a state $| \Psi \rangle$. %(we will calculate also the spectral dimension of the spatial section on 
%Gaussian states and in that case the expectation value is over the state $|c \rangle$ 
%of section (\ref{SVL})). 
Symbolically,
\begin{eqnarray}
\label{pexpect}
P(T):= \langle \hat{P_{g}}(T)\rangle = 
\int_{\Psi} Dg P(T) e^{iS(g)}.
%= \langle s | \hat{P_{g}}(T) | s \rangle
%\approx  P_{\langle s | \hat{{g}}| s \rangle}(T) = P_{\langle  g  \rangle}(T) .
\end{eqnarray}
%The third relation is not  an equality but an approximation because it is valid only in the case
%that the metric operator is diagonal on the state we are considering.  %spin-network states. 
%This is not true in general for spin-network states 
%but in our case we are interested to the scaling properties of the metric and then 
%we will not consider the non diagonal terms that are related to the non commutativity  
%(or angular part) of the metric. 
%This assumption justifies (\ref{pexpect}). 
% related to 
%
%Here $\gamma_{\mu\nu}$ denotes the microscopic metric and $S_{\rm bare}$ is the
%bare action with the gauge fixing terms and the pieces containing the ghosts
%$C$ and $\bar{C}$ included. Note that (\ref{pexpect}) does not contain any IR
%cutoff; it is the ordinary $(k=0)$ expectation value with all modes integrated
%out. In QEG the functional $S_{\rm bare}$ is given by
%the fixed point action. 
Given $P(T)$, the spectral dimension of the quantum %space or 
spacetime is defined in analogy with (\ref{dimform}):
\begin{eqnarray}
\label{specdim}
{\cal D}_{\rm s}=-2\frac{d\ln P(T)}{d\ln T}.
\end{eqnarray}
%
%Let us now evaluate the expectation value (\ref{pexpect}) using the average
%action method. 
We can formally also to replace the equation (\ref{heateq}) with 
the correspondent expectation value %on the spin-network states,
\begin{eqnarray}
\label{heateqs}
 \partial_T \langle K_{\hat{g}}(x,x';T) \rangle = 
\langle  \Delta_{\hat{g}} K_{\hat{g}}(x,x';T) \rangle. %\,\,\, \approx \,\,\,  %\nonumber \\
 %\partial_T K_{\langle \hat{g} \rangle } = 
%\Delta_{\langle \hat{g} \rangle } K_{\langle  \hat{g}  \rangle } \, ,
%\partial_T K(x,x';T)&=&\Delta(k) K(x,x';T)
\end{eqnarray}
%where $K_{\dots} = K_{\dots}(x,x';T)$. 
%The fictitious diffusion process takes place on a ``manifold''
%which, at every fixed scale $\ell \approx 1/k$, is described by a smooth Riemannian metric
%$\big<g_{\mu\nu}\big>_k$. While the situation appears to be classical at fixed
%$k$, nonclassical features emerge %in the regime with nontrivial RG running
%since at different scales different metrics apply. %there 
%The metric depends on the scale at which the spacetime structure is probed
%by a fictitious scalar field. 

{\em The Spectral Dimension in Quantum Gravity}.
In quantum gravity we define (\ref{pexpect}) the spectral dimension in the general boundary 
formalism. We introduce a gaussian state $|\psi_{\bf q} \rangle$ peaked on 
the boundary geometry
${\bf q} =(q, p)$ defined by the metric and the conjugate momentum.
We can think the boundary geometry to be the boundary of a $d$-dimensional ball.
The state is symbolically given by:
\begin{eqnarray}
\Psi_{\bf q} (s) \sim  {\rm e}^{-(s - q)^2 + i p  s}.
\label{gauss}
\end{eqnarray}
The amplitude (\ref{pexpect}) can be defined for a general spin-foam model %becomes 
\begin{eqnarray}
\hspace{-0.2cm}
\frac{\langle W | \hat{P}_g(T) | \Psi_{\rm q} \rangle}{\langle W | \Psi_{\bf q} \rangle } = \frac{\sum_{s_1, s_2} W(s_1) \, \langle s_1 | \hat{P}_g |s_2 \rangle \, \psi_{\bf q} (s_2)}
{\sum_s W(s) \Psi_{\bf q}(s)} .
\label{pmedgb}
\end{eqnarray}
Where $W(s)$ codifies the spin-foam dynamics \cite{DO}. For the purpose of
the paper we will consider the PR model (TV model); the vertex amplitude is 
encoded in the $\{6 j\}$-symbol, $W(s) \propto \{6 j\}$ ($W(s) \propto \{6 j\}_q$ for TV
and $q$, the quantum deformation of the $SU(2)$ group, is related to the cosmological constant 
$\Lambda$ by $q=\exp(2 i \sqrt{\Lambda}l_P$).
Since we are interested in the scaling of the Laplacian to 
analyze the fractal properties of the space-time, we can approximate the 
metric in the Laplacian 
with the inverse %of the square 
of the $SU(2)$ Casimir operator. 
%Using this simplification we define the operator $\hat{P}_g(T)$ in the following way,
%\begin{eqnarray} \hspace{-0.2cm} 
%\hat{P}_g(T) = V^{-1} \, {\rm Tr} \, \big(  {\rm e}^{T \hat{\Delta}_g} \big)
 %\rightarrow \hat{\mathbb{P}}(T) = V^{-1} {\rm Tr} \, \big( {\rm e}^{T \, \widehat{\frac{\Delta_0}{C^2}} }\big).
%\label{PTS}
%\end{eqnarray} 
%Where $\Delta_0$ is the Laplacian in the infrared limit.
We recall that in $3d$ quantum gravity the Casimir operator is related to the
length spectrum of a link $e$ in the simplycial decomposition by the relation \cite{LoopOld}
\begin{eqnarray}
L_e^2 = l_P^2 C^2(j_e) = l_P^2[ \, j_e(j_e+1) + c \, ],
\label{length}
\end{eqnarray}
where the constant is chosen to be $c=1/4$ in line with \cite{speziale}.
%We introduce a Diff-invariant scale defined by $\ell :=j_e l_P$. 
In $3d$ gravity we approximate the $3$-ball with a single tetrahedron and
the boundary $S^2$ sphere by the surface of the tetrahedron
given by the six triangles.
We consider fixed four of the six representations ($j$) and we call the other two
free representations by
$j_e$ ($e =1,2$).
Following the ideas and notation above 
we define the operator $\hat{P}_g(T)$ in the following way,
\begin{eqnarray} \hspace{-0.2cm} 
%\hat{P}_g(T) %= V^{-1} \, {\rm Tr} \, \big(  {\rm e}^{T \hat{\Delta}_g} \big)\rightarrow 
 \hat{ P}_{j_e}(T) := V^{-1} {\rm Tr} \,\, {\rm e}^{T \frac{C^2_0}{C^2_e}\Delta_0} 
:= V^{-1} {\rm Tr} \,\, \hat{{\mathcal O}}_e.
\label{PTS}
\end{eqnarray} 
Where $\Delta_0$ is the Laplacian at a lower infrared scale, 
%
%The operator $\hat{{\mathbb P}}(T)$ defined in (\ref{PTS}) is defined in one
%of the two links $j_e$ and then it becomes 
%\begin{eqnarray}
%\hat{{\mathbb P}}_{j_e}(T) :=V^{-1} {\rm Tr} \Big( {\rm e}^{T \widehat{\frac{\Delta_0}{C^2(j_e)}}} \Big)
%:= V^{-1} {\rm Tr} \big(\hat{{\mathcal O}}_e\big),
%\label{Pf}
%\end{eqnarray}  
$j_e$ is fixed (for example) to $j_e = j_1$ and 
\begin{eqnarray}
\hat{{\mathcal O}}_e :=  {\rm e}^{T \frac{C^2_0}{C^2_e}\Delta_0}.
\label{oop}
\end{eqnarray}
The boundary state in the notation above is  
\begin{eqnarray}
\hspace{-0.2cm}
\Psi_{j}(j_e) = {\mathcal N}^{-1} \, 
 {\rm e}^{- \frac{2}{3 j} \sum_{e}^2 (j_e - j)^2 + i \theta \sum_{e}^2(j_e+1/2)}.
\label{state0}
\end{eqnarray} 
where ${\mathcal N}$ is a normalization factor. The dihedral angles $\theta = \arccos (-1/3)$
define the boundary extrinsic geometry for an equilateral tetrahedron.
Now we have all the ingredients to calculate the expectation value (\ref{pmedgb})
using (\ref{oop}) and (\ref{state0}), In particular, since the geometry appears only in the
operator $\hat{{\mathcal O}}_e$, we can calculate the expectation value of this operator,
\begin{eqnarray}
&& \hspace{-0.7cm} 
\eta \, \langle W| \hat{{\mathcal O}}_{j_1} |\Psi_j \rangle  =
\eta \hspace{-0.1cm}
\sum_{j_{1,2}=0}^{2j} W(j_1, j_2, j) {\mathcal O}_{j_1} \Psi_{j}(j_{1, 2})%}{\langle W| \Psi_j \rangle}
\nonumber \\
&& \hspace{-1cm}=\eta 
\hspace{-0.1cm}
\sum_{j_{1,2} =0}^{2 j} 
\prod_{e=1}^6 (2 j_e+1) \{6 j\} \, {\rm e}^{- T \frac{C_0^2 |\Delta_0|}{[ j_1 (j_1 +1) +c ]}} 
\,   \Psi_j(j_{1,2} ),
%\Psi_j(j_1, j_2) 
%\nonumber \\
%&&
\label{ampli}
\end{eqnarray}
where we introduced the following notation for the normalization, 
$\eta^{-1} := \langle W| \Psi_j \rangle$. We also replaced the Laplacian $\Delta_0$
with $- |\Delta_0|$. 
Before to calculate the amplitude (\ref{ampli}) we replace $\Delta_0$ with
$|\Delta_0| \propto 1/j^2$, this assumption will be clear later in the paper.

\begin{figure}
 \begin{center}
 \hspace{-0.5cm}\includegraphics[height=3.5cm]{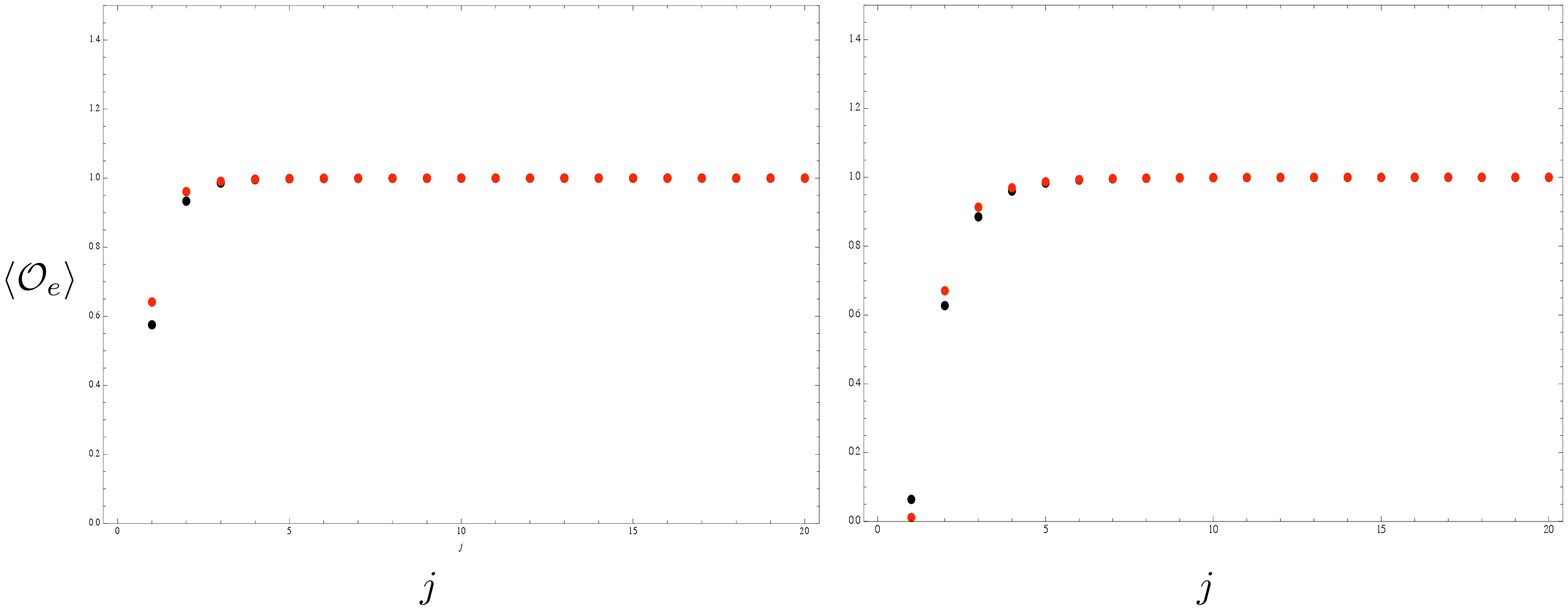}
    \end{center}
    \vspace{-0.3cm}
  \caption{\label{PlotDot1} 
 This is the plot of the modulus of the expectation value (\ref{ampli}) 
 $|\eta \, \langle W| \hat{{\mathcal O}}_{j_1} |\Psi_j \rangle|$  (black dots)
 compared with the exponential $\exp(-T C_0^2 |\Delta_0|/[j (j+1) +c])$ in the case $c=1/4$
 (red dots). The expectation value is calculated for $T=1$ and $T=10$,
 $0 \leqslant j_1, j_2 \leqslant 2 j$. }
  \end{figure}

\begin{figure}
% \begin{center}
 \includegraphics[height=3.2cm]{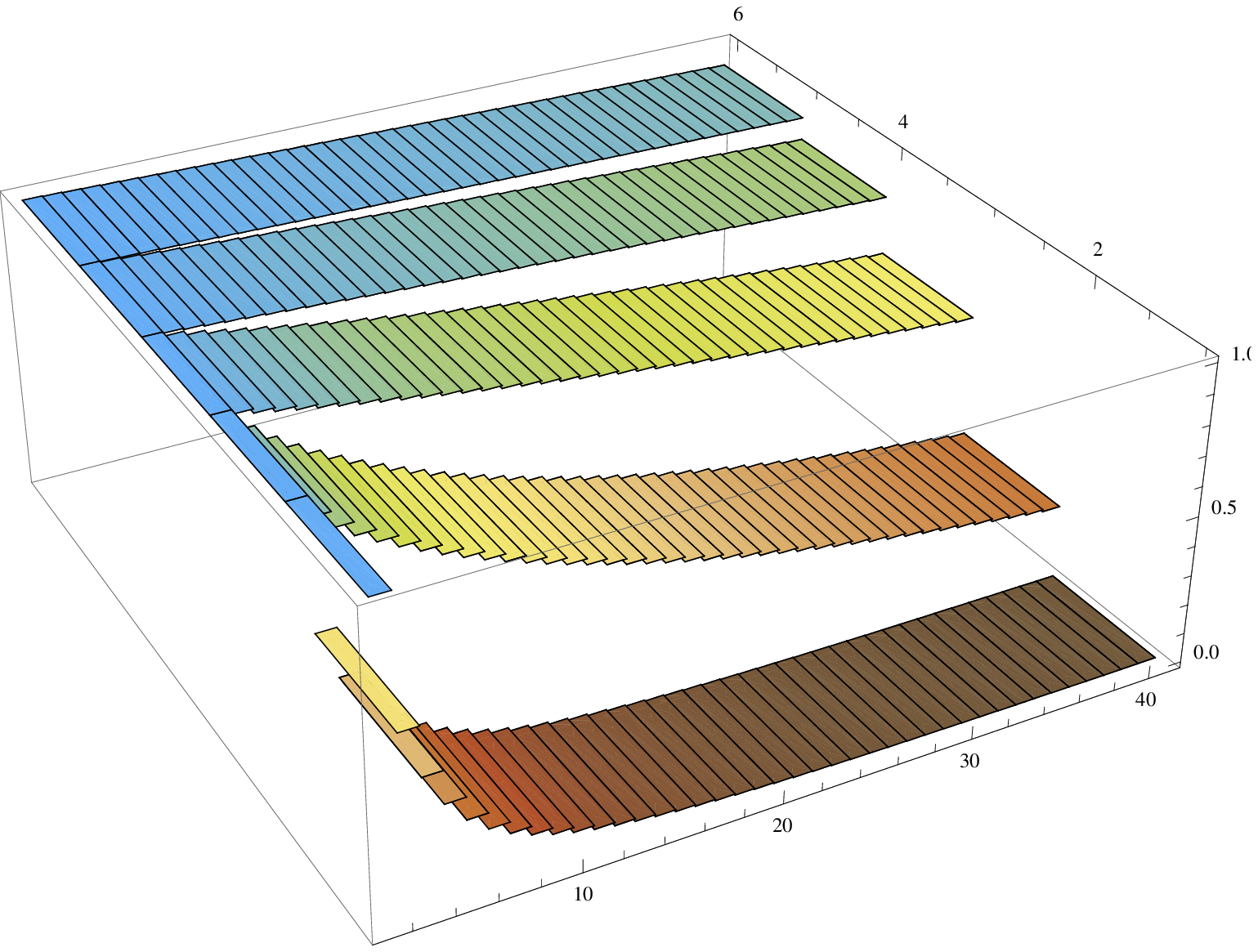}
 \includegraphics[height=3.2cm]{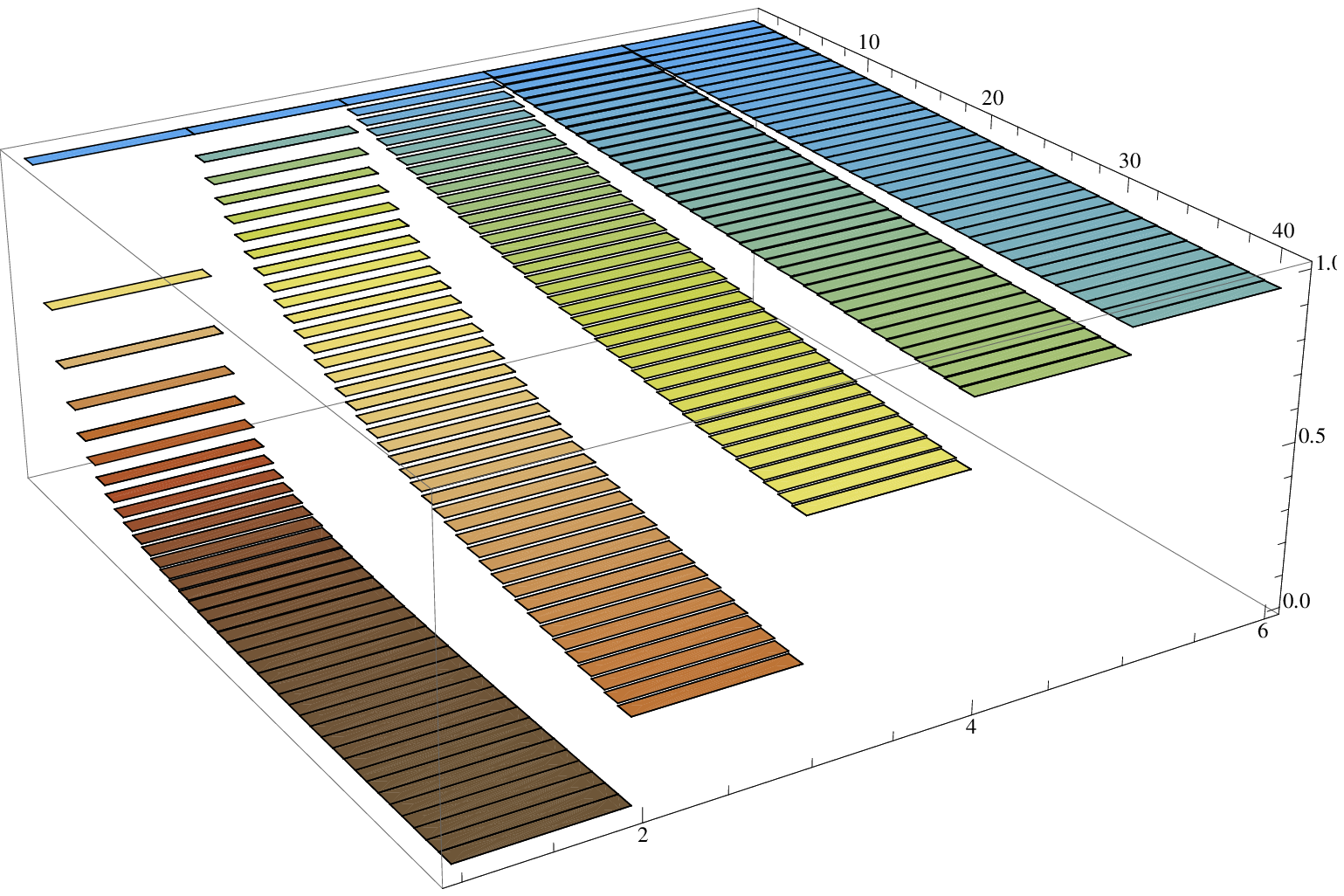}
 %   \end{center}
    \vspace{-0.3cm}
  \caption{\label{OjT} 
  Plot of the amplitude (\ref{ampli}) for $1 \lesssim j \lesssim 6$ 
  and $1 \lesssim T C_0^2  \lesssim 40$. 
  }
  \end{figure}

\begin{figure}
% \begin{center}
 \includegraphics[height=3.4cm]{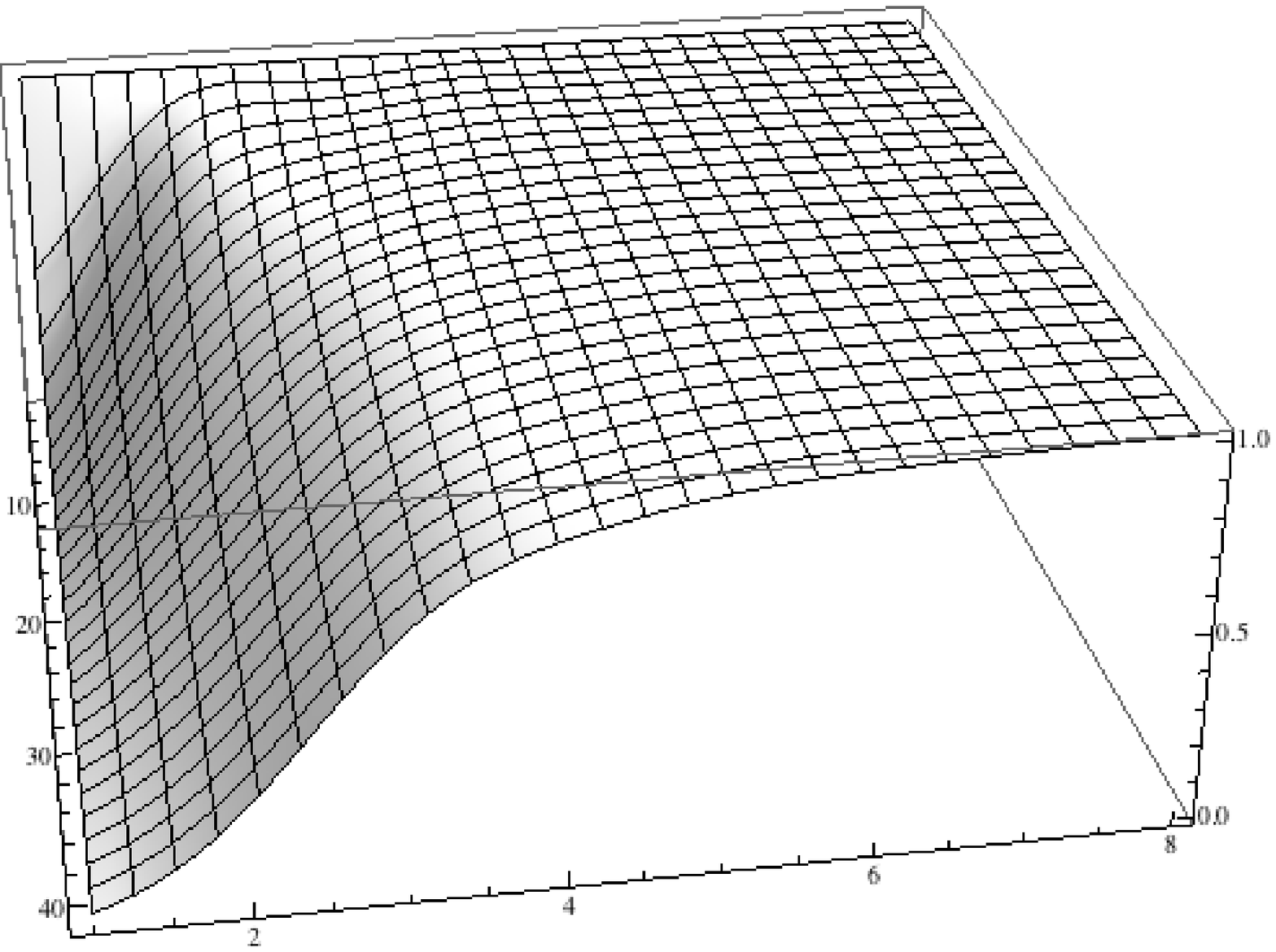}
 \includegraphics[height=3.6cm]{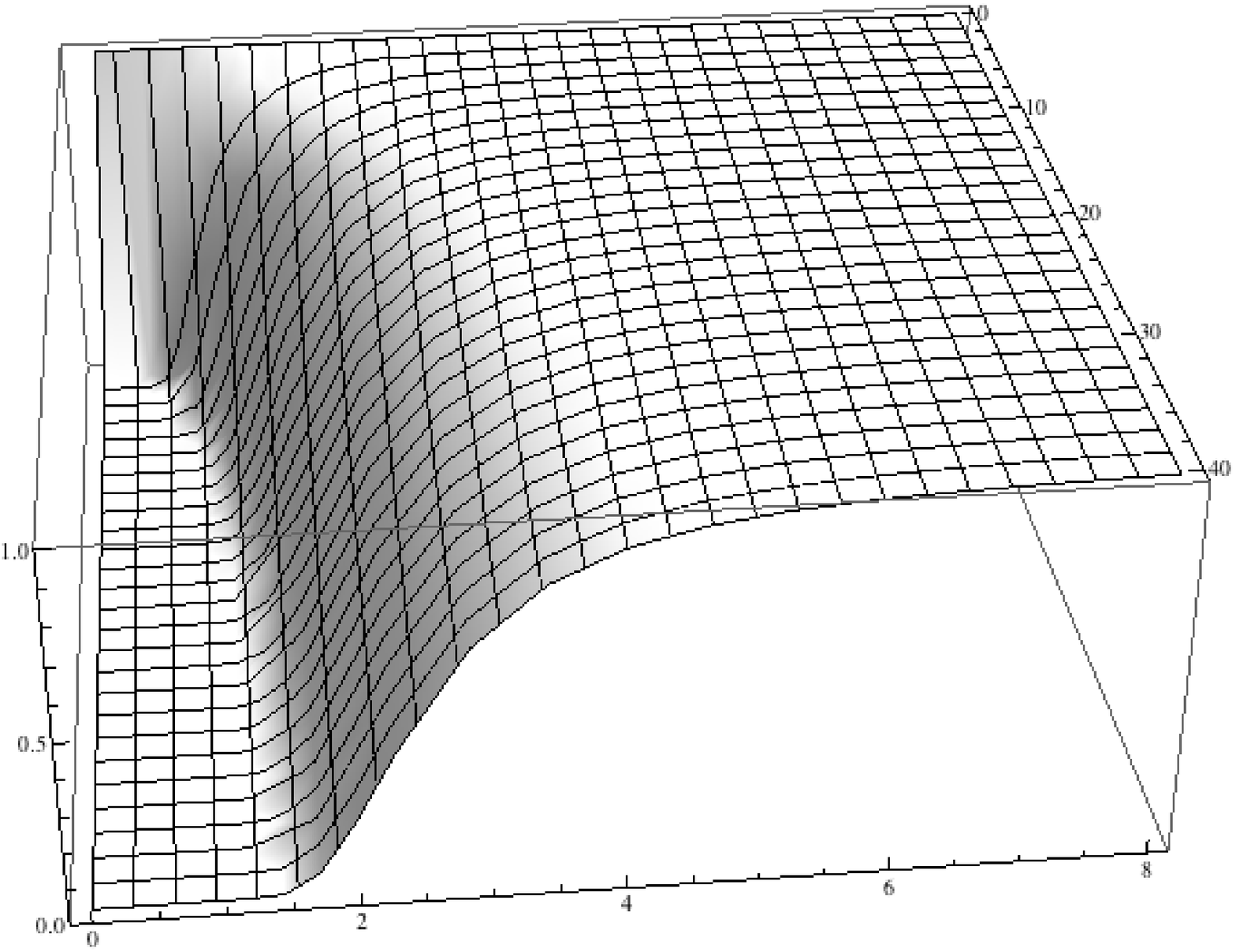}
 %   \end{center}
   % \vspace{-0.3cm}
  \caption{\label{OjT2} 
  Plot of the amplitude (\ref{ampli}) for $1 \lesssim j \lesssim 8$ 
  and $1 \lesssim X  \lesssim 40$ on the left and of the function 
  $\exp(-X |\Delta_0|/[j (j+1) +1/4])$ on the right, where $X= T C_0^2$
  This plots show there is good agreement for $j \gtrsim 4$ . 
  }
  \end{figure}

\begin{figure}
% \begin{center}
 \includegraphics[height=4.2cm]{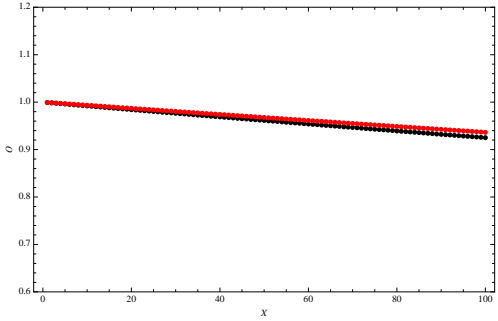}
 %\includegraphics[height=3.2cm]{PlotOjT2.eps}
 %   \end{center}
    \vspace{-0.3cm}
  \caption{\label{Ojfix} 
  Plot of the amplitude (\ref{ampli}) for $j = 6$ as function of $X= T C_0^2$ on the left (black points)
  and of the function $\exp(-X |\Delta_0|/[j (j+1) +1/4])$ $1 \lesssim T C_0^2  \lesssim 40$. 
  }
  \end{figure}

The result of the calculation (\ref{ampli}) is given in Fig.\ref{PlotDot1} and compared 
with the exponential $\exp[- T C_0^2|\Delta_0|/(j (j+1) +c)]$ in the case $c=1/4$,
$C_0^2 =1$ and $|\Delta_0| \propto 1/j^2$. The plots in Fig.\ref{PlotDot1}
are for $T=1$ and $T=10$. 
We can observe a perfect agreement for $j \gtrsim 4$.
This agreement is supported by the plots in Fig.\ref{OjT} and Fig.\ref{OjT2} 
where the amplitude (\ref{ampli}) on the left and the function 
$\exp(-X |\Delta_0|/[j (j+1) +1/4])$ ($X= T C_0^2$) on the right 
coincide for $j \gtrsim 4$. In Fig.\ref{Ojfix} we plotted a section 
of (\ref{ampli}) for $j=6$ and $X \in [0, 100]$. This section coincides 
with the function $\exp(-X |\Delta_0|/[j (j+1) +1/4])$ evaluated on $j=6$.

\begin{figure}
% \begin{center}
 \includegraphics[height=3.2cm]{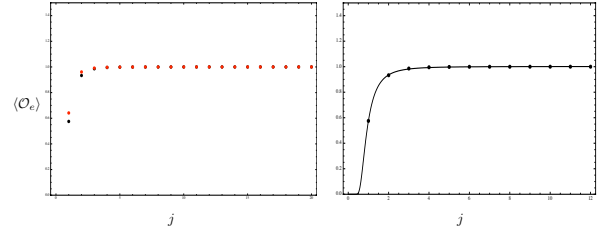}
 %\includegraphics[height=2.5cm]{FIT23.eps}
 %   \end{center}
    \vspace{-0.3cm}
  \caption{\label{Fit} 
  The plot on the left represents the points obtained from the evaluation 
  of the amplitude(\ref{ampli}) for $1 \lesssim j \lesssim 20$ (black points). 
  The red points refer to the function $\exp(-T C_0^2\Delta_0/[ \,j (j+1) +c \,])$
  for $C_0^2=1$ (it is an irrelevant constant) and $T=1$;
  it is evident that for small values of the representation $j$ the two function 
  are different.
  The plot on the right
  represents an interpolation of the black points in the picture on the left
  for $1 \lesssim j \lesssim 12$. 
  }
  \end{figure}

In the range $1 \lesssim j \lesssim 12$ we have interpolated the exact result (\ref{ampli}) numerically 
and obtained a different exponential form of the amplitude. The points data and the fit 
are given in Fig.\ref{Fit}. The points are fitted by the function $a \exp(b/j^{\alpha})$, where
$a \approx 1.00$, $b \approx 0.55$ and $\alpha \approx 3.03$ for $T=1$.
Recalling that $\Delta \propto 1/j^2$ we conclude that at the Planck scale,
\begin{eqnarray}
\eta \, \langle W| \hat{{\mathcal O}}_{j_1} |\Psi_j \rangle  \approx {\rm e}^{-T \, 0.55 \, \frac{|\Delta_0|}{j^{1.03}}}.
\label{Hscale}
\end{eqnarray}
%We can not approximate the data for $1 \lesssim j \lesssim 12$ with a continuum function 
%because the continuum fit  is a good fit for $j\gtrsim 12$.
We will use this result to calculate the spectral dimension at the Planck scale then
for $T\approx 1$ in Planck units, this is the reason why we fixed $T=1$ 
in the expectation value.
We can reproduce the behavior of (\ref{ampli}) for $j\gtrsim 4$ 
(in Fig.\ref{Fit} the function $\exp(-T C_0^2\Delta_0/[ \,j (j+1) +c \,])$ coincides perfectly with 
the exact expectation value (\ref{ampli}) from $j\gtrsim 4$) 
also analytically 
using the asymptotic large $j$ limit of the $\{6 j\}$ symbol. For large $j$ we have: 
$\{6 j\} \propto \exp( i S_R(j_e)+ i \pi/4) + c.c$. Using this property of the symbol and
replacing the sum in (\ref{ampli}) with an integral on $\delta j_{1,2} :=j_{1,2} - j$ (for $j\gg1$)
we obtain $\exp(T C_0^2 \Delta_0/[j (j+1) +c])$.

What we learnt from the explicit calculation of (\ref{pmedgb}) can be summarized as follows,
%we can replace 
\begin{eqnarray}
\langle \hat{{\mathcal O}}_e \rangle  \approx \left\{ \begin{array}{ll} 
          e^{- T \frac{C_0^2}{j (j+1) +c} \Delta_0} & {\rm for} \,\,\, j\gg1  \,\,\, (j \gtrsim 4), \\ 
         a \, e^{- T \frac{b \, C_0^2}{j^{\alpha}} \Delta_0}  & {\rm for} \,\,\, j \approx 1 \,\,\, (1\lesssim j \lesssim 4).
        \end{array} \right. 
\label{Flimits}
\end{eqnarray}
Where $\alpha \approx 1.03$.
We introduce a Diff-invariant scale defined by $\ell :=j_e l_P$
The result of the (\ref{pmedgb}) can be summarized in the scaling property of 
the Laplacian operator 
%the metric 
with the scale $\ell$ (or with the energy scale $k \approx 1/\ell$),
\begin{eqnarray}
\Delta_j  \approx  \left\{ \begin{array}{ll} 
         \frac{C_{j_0}^2}{j (j+1) +c} \Delta_{j_0} &{\rm for} \,\,\, j\gg1  \,\,\, (j \gtrsim 4), 
         \vspace{0.1cm}
         \\
                  \frac{c^{\prime }}{j^{\alpha}} \Delta_{j_0^{\prime}}  & 
         {\rm for} \,\,\, j \approx 1 \,\,\, (1\lesssim j \lesssim 4).
        \end{array} \right. 
\label{deltalimits}
\end{eqnarray}
Where we introduced the infrared scales $0\rightarrow j_0, j_0^{\prime}$
$j_0 \gg 1$, $j_0^{\prime} \gtrsim 4$), $c^{\prime} = b C_0^2$ and, 
by definition, $C^2_{j_0} = j_0(j_0+1) +c$. % and $C^2_{j_0} = j_0(j_0+1) +c$
%The behavior of the Laplacian for $ 1\lesssim j \lesssim 6$ is correct only for 

We denote the scaling of the Laplacian operator suggested by (\ref{deltalimits}) by a general function 
in the momentum space. We introduce here a physical input to put the momentum $k$ in our 
analysis.
If we want to observe the space-time with a microscope of resolution $l=l_P j$ 
(the infrared length is $\ell_0 := l_P j_0$) 
we must use a (fictitious) probing scalar field of momentum $k \sim 1/l$. 
The scaling property of the Laplacian in terms of $k$ can be obtained 
by replacing: $l\sim 1/k$, $l_0\sim 1/k_0$ and $l_P\sim 1/E_P$,
where $k_0$ is an infrared energy cutoff and $E_P$ is the Planck energy.
%The nonclassical features are encoded in the properties of the diffusion
%operator. 
We define the covariant Laplacian at the scale $k$ introducing the
function ${\mathbb S}_k$,
\begin{eqnarray}
\label{opscale0}
\Delta_k =\mathbb{S}_k (k,k_0)\,\Delta_{k_0}.
\end{eqnarray}
It is straightforward to derive the scaling function from (\ref{deltalimits}) and using the arguments 
above
\begin{eqnarray}
{\mathbb S}_k(k, k_0)  \approx  \left\{ \begin{array}{ll} \hspace{-0.1cm}
         \frac{k^2[ \, E_P(E_P + k_0) + c k_0^2 \, ]}{k_0^2 [ \, E_P(E_P +k) +c k^2 \,]}  + 1& 
    \hspace{0cm}     {\rm for} \,\,\,  k \lesssim \frac{E_P}{4}, 
         \vspace{0.1cm}
         \\\hspace{-0.1cm}
                  c^{\prime } \, k^{\alpha}  & 
  \hspace{-2.5cm}       {\rm for} \,\,\, \frac{E_P}{4} \lesssim k \lesssim E_P.
        \end{array} \right. 
\label{Sklimits}
\end{eqnarray}
We added a factor one in the infrared limit to facilitate the spectral dimension
calculations.
The scaling function ${\mathbb S}_k(k, k_0)$ represents also , using the definition of the Laplacian,
 the scaling of the inverse of the metric, 
 $\langle g^{\mu \nu} \rangle_k = {\mathbb S}_k (k, k_0)\langle g^{\mu \nu} \rangle_{k_0}$.

We suppose that the diffusion process involves (approximately) only a small interval of 
scales near $k$ %over which $\bar{\lambda}_k$ does not change much the 
then the corresponding heat kernel contains the $\Delta_k$ for this specific and fixed
value of the momentum scale $k$.
% quantization 
%\begin{eqnarray}
%\label{heateqk}
%\partial_T \langle s|K_{\hat{g}}(x,x';T) |s \rangle &=& 
%\langle s| \Delta_{\hat{g}}(k) K_{\hat{g}}(x,x';T)| s \rangle, \\
%\partial_T K_{\langle s |\hat{g}| s \rangle }(x,x';T) &=& 
%\Delta_{\langle s |\hat{g}| s \rangle }(k) K_{\langle s | \hat{g} | s \rangle }(x,x';T), \\
%\partial_T K(x,x';T)&=&\Delta(k) K(x,x';T), % \,\,\, \approx \,\,\, \partial_T K(x,x';T)&=&\Delta(k) K(x,x';T),
%\end{eqnarray}
%
%The equation (\ref{heateqk}) is exactly (\ref{heateqs}) where we have suppressed 
%the index $\langle \hat{g} \rangle$, and introduced the Laplacian at the scale $k$ in terms 
%of the Laplacian at the scale $k_0$.
\noindent Denoting the eigenvalues of $-\Delta_{k_0}$ by ${E}_n$
%$\Delta(k_0) |E_n \rangle = - E_n |E_n \rangle$,
and the corresponding eigenfunctions by $\phi_n(x) = \langle x| E_n \rangle$, 
we have the following eigenvalue equation for the Laplacian %at the scale $k_0$,
\begin{eqnarray}
&& \hspace{1.5cm}\Delta_{k_0} |E_n \rangle = - E_n |E_n \rangle, \nonumber \\
&& \hspace{-0.8cm} \langle E_n| E_m \rangle = \int d^4x'\,\sqrt{g_0(x')} \, \phi^*_n(x^{\prime}) \phi_{n}(x^{\prime}) = \delta_{n, m}.
\label{Deltak0}
\end{eqnarray}
Using (\ref{Deltak0}) and the definition (\ref{EK})
we can calculate explicitly the heat kernel %. The result is 
 %the equation (\ref{heateqk}) is solved by
%
%\begin{eqnarray}
%\hspace{-0.5cm} K_k(x,x';T)&=&\sum\limits_n\phi_n(x)\,\phi^*_n(x')\, e^{-{\mathbb S}_k(k, k_0)\, E_n
%\,T}.
%\label{kernexp}
%\end{eqnarray}
%
%(see the footnote)\footnote{
%\fbox{\parbox{3.0cm}{{\em Proof 1} of (\ref{kernexp})}} \\
%
%It is very simply to show 
%\begin{center}
%\fbox{\parbox{15.7cm}{
%\fbox{\parbox{3.0cm}{{\em Proof 2} of (\ref{kernexp})}} \\
%{\em Proof of (\ref{kernexp}).}
%We want to calculate 
$K_{k}(x,x';T)  = \langle x^{\prime} | \langle \hat{{\mathcal O}}_e \rangle | x \rangle$. 
%using the definition given at the beginning of the
%paper. 
By using (\ref{Flimits}), (\ref{deltalimits}), (\ref{opscale0}) and (\ref{Deltak0}) we have %the solution of (\ref{heateqk}) is: 
\begin{eqnarray}
&& K_{k}(x,x';T) 
= \langle x^{\prime}| e^{T\Delta_{k}} | x \rangle \nonumber \\
%&& \hspace{0.0cm} = \sum_n \sum_m \langle x^{\prime}| E_{n^{\prime}} \rangle \langle E_{n^{\prime}}| 
%e^{T\Delta_{k}} | E_{n} \rangle \langle E_{n}| x \rangle \nonumber \\
%&& \hspace{0.0cm} = \sum_n \sum_m \phi^*_n(x^{\prime}) \langle E_{n^{\prime}}| 
%e^{T \, {\mathbb S}_k (k, k_0) \Delta(k_0)} | E_{n} \rangle \phi_{n}(x) \nonumber \\
%&& = \sum_n \sum_m \phi^*_n(x^{\prime}) \langle E_{n^{\prime}}| 
%e^{ - T \, {\mathbb S}_k (k, k_0) E_n} | E_{n} \rangle \phi_{n}(x) \nonumber 
%\end{eqnarray}
%\begin{eqnarray}
%&& = \sum_n \sum_m \phi^*_n(x^{\prime}) \phi_{n}(x) \, \delta_{E_{n^{\prime}}, E_n} \,  
%e^{- T \, { \mathbb S}_k (k, k_0) E_n}\nonumber \\
&& \hspace{0.05cm}
 =\sum_n  \phi^*_n(x^{\prime}) \phi_{n}(x) \, 
e^{- T \, { \mathbb S}_k (k, k_0) E_n}.
\label{Ktra}
%\label{kernexp}
\end{eqnarray}
%\fbox{\parbox{4.0cm}{{\em End proof} of (\ref{kernexp})}}
%   }}
 %  \end{center}
 %
 % \begin{center}
 % \fbox{\parbox{15.7cm}{
%We have just shown that the LHS and RHS of (\ref{kernexp}) are equal. 
%}}
%\end{center}
%
%Here we introduced the convenient notation $F(k^2)\equiv\bar{\lambda}_k/
%\bar{\lambda}_{k_0}$. 
%
From the knowledge of the propagation kernel (\ref{Ktra}) we can time-evolve any
initial probability distribution $p(x;0)$ according to
$p(x;T)=\int d^4x'\,\sqrt{g_0(x')}\,K(x,x';T)\,p(x';0)$, where $g_0$ the 
determinant of $\big<g_{\mu\nu}\big>_{k_0}$. If the initial distribution has 
an eigenfunction expansion of the form $p(x;0)=\sum_n C_n\,\phi_n(x)$ we
obtain for arbitrary $x$, 
\begin{eqnarray}
\label{probexp}
&& p(x;T)= \int d^4x'\,\sqrt{g_0(x')}\,K(x,x';T)\,p(x';0) = \nonumber \\
%&& = \sum_n \sum_m \, \int d^4x'\,\sqrt{g_0(x')}\,
 % \phi^*_n(x^{\prime}) \phi_{n}(x) \nonumber \\
%&& \hspace{1.8cm} e^{- T \, { \mathbb S}_k (k, k_0) E_n}  \,
% C_m\,\phi_m(x^{\prime})
%\nonumber  \\
&& =\sum_n C_n\,\phi_n(x)\, e^{-{\mathbb S} (k, k_0)\, E_n\,T}
\end{eqnarray}
%
%\vspace{-0.5cm}
%
%\fbox{\parbox{15.7cm}{
%\fbox{\parbox{2.6cm}{{\em Proof} of (\ref{probexp0})}} 
%
%To pass 
where we used the wave function normalization (\ref{Deltak0}).
%}}
%\vspace{0.5cm}\\
%\fbox{\parbox{3.5cm}{{\em End proof} of (\ref{probexp0})}} 
%{\bf End Proof.}
If the $C_n$'s are significantly different from zero only for a single
eigenvalue ${E}_n$, we are dealing with a single-scale problem and 
%. In the
%usual spirit of effective field theories we would 
then we can identify 
$k^2={E}_n$. %as the relevant scale at which the running couplings are to be evaluated.
However, in general the $C_n$'s are different from zero over a wide range of
eigenvalues. In this case we face a multiscale problem where different modes
$\phi_n$ probe the spacetime on different length scales.

If $\Delta(k_0)$ is the Laplacian on the %corresponds to 
flat space, the eigenfunctions $\phi_n
\equiv\phi_p$ are plane waves with momentum $p^\mu$, and they resolve
structures on a length scale $\ell$ of order $1/|p|$. Hence, in terms of the
eigenvalue $E_n\equiv{ E}_p=p^2$ the resolution is $\ell \approx
1/\sqrt{{E}_n}$. This suggests that when the manifold is probed by a
mode with eigenvalue ${E}_n$ it ``sees'' the metric 
$\big<g_{\mu\nu}\big>_k$ for the scale $k=\sqrt{{E}_n}$. Actually the
identification $k=\sqrt{{E}_n}$ is correct also for a curved spacetime
because the parameter $k$ just identifies the scale we are probing.
%independently from the curvature. 
%if we consider a short time interval $T$ compared to the scale $\ell$.
%since,
%in the construction of $\Gamma_k$, the parameter $k$ is introduced precisely
%as a cutoff in the spectrum of the covariant Laplacian. 
Therefore we can conclude that under the spectral sum of (\ref{probexp}) we must
use the scale $k^2={E}_n$ which depends explicitly on the resolving power
of the corresponding mode. 
%Likewise, i
In eq. (\ref{probexp}), ${\mathbb S}_k(k, k_0)$ can be
interpreted as ${\mathbb S}({E}_n)$. %${\mathbb S}_{E_n}({E}_n)$. 
Thus we obtain the traced propagation kernel,
%
%
%

%strong RG effects any more.\\
%\vspace{0.0cm}
%\hspace{-0.415cm}
%\begin{center}
%\fbox{\parbox{15.7cm}{
%\fbox{\parbox{2.6cm}{{\em Proof} of (\ref{trpropk})}} 
%{\bf Proof of (\ref{trpropk})}.
\begin{eqnarray}
%&& 
\hspace{-0.2cm} P(T) =   %\langle V^{-1} \, \int \sqrt{g} \, d^d x \,  K_{g}(x, x;T) \rangle \nonumber \\
% && \hspace{1.2cm} 
 %= \langle   \left( V^{-1} \, \int \sqrt{g_k} \, d^d x \, 
%\langle x | e^{T \Delta_k} | x \rangle \right)  \rangle \nonumber \\
%
%&& \hspace{0.0cm} = V^{-1}(\langle g \rangle_k)  \, \int \sqrt{\langle g \rangle_k} \, d^d x \, 
%langle x | e^{T \Delta_{k}} | x \rangle  \nonumber\\
%&& \hspace{0cm} = 
%\Big(\sum_n   \, \int \sqrt{{\mathbb  S}^{-d}(k, k_0)} \sqrt{\langle g \rangle}_{k_0} \, d^d x \, 
%\phi^*_n(x) \phi_{n}(x) \nonumber \\
%&& \hspace{0.5cm} e^{- T \, { \mathbb S}_k (k, k_0) E_n} \Big)/
%\Big(\int \sqrt{{\mathbb  S}^{-d}(k, k_0)} \sqrt{\langle g \rangle}_{k_0}
%d^d x \Big)
%\nonumber \\
%\end{eqnarray}
%\begin{eqnarray}
%
%
%&& \hspace{0cm}
 %= 
%\frac{\sum_n   e^{- T \, { \mathbb S}_k (k, k_0) E_n}}{\int \sqrt{\langle g \rangle}_{k_0} 
%d^d x} \,\,\,\,\,\,  
%\underrightarrow{{k^2 \approx E_n}} \,\,\,\,\,\,  
\sum_n \, \frac{e^{- T \, { \mathbb S} (E_n) E_n}}{V_{\langle g \rangle_{k_0}}}  %\nonumber \\  
%&& =  V^{-1}\;\sum_n e^{-{\mathbb S}({ E}_n)\,{E}_n\,T} \nonumber \\
%&& 
= V^{-1}{\rm Tr} \left(e^{ {\mathbb S}(-\Delta_{k_0} )\,\Delta_{k_0}\,T}\right)
\hspace{-0.1cm}. 
 \label{proofPT}
\end{eqnarray}
%The dependence by $\langle g \rangle_k$ in the second line is obtained 
%evaluating the volume $V$ at the scale $k$.
%We have used (\ref{kernexp}) and (\ref{opscale0}) from the third to the forth line,
%(\ref{norm}) in the last line. 
It is convenient to choose $k_0$ as a macroscopic scale in a regime where there are not strong quantum gravity effects. 
%\\ {\bf End Proof.} % of (\ref{trpropk})}\\
%\fbox{\parbox{3.5cm}{{\em End proof} of (\ref{trpropk})}}
%The average on spin-network state is: 
%\begin{eqnarray}
%\langle s | g_{\
%\end{eqnarray}
%}}
%\end{center}
%\vspace{0.5cm}
%Furthermore, let us assume for a moment that at $k_0$ the cosmological
%constant is tiny, $\bar{\lambda}_{k_0}\approx 0$, so 

We assume for a moment 
that $\big<g_{\mu\nu}
\big>_{k_0}$ is an approximately flat metric. In this case the trace in eq.
(\ref{proofPT}) is easily evaluated in a plane wave basis:
\begin{eqnarray}
&& \hspace{-0.2cm} P(T)  %:= P_{\langle g \rangle_k} (T) \nonumber \\
% = V^{-1}(\langle g \rangle_k)  \, \int \sqrt{\langle g \rangle_k} \, d^d x \, 
%\langle x | e^{T \Delta_{k}} | x \rangle  \nonumber\\
%&& \hspace{-0.2cm}= V^{-1}(\langle g \rangle_k)  \, \int \int \sqrt{\langle g \rangle_k} \, d^d x \, \frac{d^d p}{(2 \pi)^d} \, 
%\langle x | e^{T \Delta_{k}} | p \rangle  \langle p | x \rangle  \nonumber\\
%&&  \hspace{-0.2cm}= {\mathbb S}(k, k_0)^{d/2} \, V^{-1}(\langle g \rangle_{k_0})  \, \int \int 
%{\mathbb S}(k, k_0)^{-d/2}
%\sqrt{\langle g \rangle_{k_0}} \, d^d x \nonumber \\
%&& \hspace{3cm} \frac{d^d p}{(2 \pi)^d} \, 
%\langle x | p \rangle \, e^{- T \, {\mathbb S} (k, k_0) p^2}  \langle p | x \rangle  \nonumber\\
%&& \hspace{-0.2cm}= \frac{\int 
%\sqrt{\langle g \rangle_{k_0}} \, d^d x \, \int \frac{d^d p}{(2 \pi)^d} \, 
%e^{i p x} \, e^{- T \, {\mathbb S} (k, k_0) p^2}  \, e^{- i p x}}{V(\langle g \rangle_{k_0})}  \nonumber\\
%&& \hspace{-1.0cm}= \int \frac{d^d p}{(2 \pi)^d} \, e^{- T \, {\mathbb S} (k, k_0) p^2}  % \nonumber \\
%\,\,\,\,\,\, 
%&& 
%\hspace{0.0cm}\underrightarrow{{k^2 \approx p^2}} \,\,\,\,\,\,  
= \int \frac{d^d p}{(2 \pi)^d} \, e^{- T \, {\mathbb S} (p) p^2} \hspace{-0.0cm}.
% \label{dimPOP}
\label{trplane}
\end{eqnarray}
%
%\begin{eqnarray}
%\label{trplane}
%P(T)=\int\frac{d^4p}{(2\pi)^4}\, e^{-p^2\, {\mathbb S}(p)\,T}.
%\end{eqnarray}
%
where we used the flat metric $\langle g_{\mu \nu} \rangle_{k_0} = \delta_{\mu \nu}$ and 
$\Delta_{k_0} | p \rangle = - p^2 | p \rangle$.

The dependence from $T$ in (\ref{trplane}) determines the fractal dimensionality of
spacetime via (\ref{specdim}). In the limits $T\rightarrow\infty$ and 
$T\rightarrow 0$ where we are probing very large and small distances,
respectively, we obtain the dimensionalities corresponding to the largest
and smallest length scales possible. The limits $T\rightarrow\infty$ and 
$T\rightarrow 0$ of $P(T)$ are determined by the behaviour of ${\mathbb S}(p)$ %\equiv
%\bar{\lambda}(k=\sqrt{p^2})/\bar{\lambda}_{k_0}$ 
for $p\rightarrow 0$ and
$p\rightarrow\infty$, respectively.\\
%The quantum gravity effects stop below some scale energy 
%that we denoted by $k_0$ and 
%we have ${\mathbb S}(p\rightarrow 0)=1$. In this case (\ref{trplane}) yields
%$P(T)\propto 1/T^2$, and we conclude that the macroscopic spectral dimension
%is ${\cal D}_{\rm s}=4$. %In the next section we apply the introduced ideas 
%to the spatial section in LQG and to the space-time in the covariant spin-foam 
%formulation of quantum gravity.
%%%%%%
%In the fixed point regime we have $\bar{\lambda}_k\propto k^2$, and therefore
%$F(p^2)\propto p^2$. As a result, the exponent in (\ref{trplane}) is 
%proportional to $p^4$ now. This implies the $T\rightarrow 0-$behavior
%$P(T)\propto 1/T$. It corresponds to the spectral dimension 
%${\cal D}_{\rm s}=2$.
%This result holds for all RG trajectories since only the fixed point 
%properties were used. In particular it is independent of $\bar{\lambda}_{k_0}$
%on macroscopic scales. 
%The result we will find about %the hight energy spectral dimension 
%are of general character. 
The above assumption that $\big<g_{\mu\nu}\big>_{k_0}$ 
is flat was not necessary to obtain the spectral dimension at 
any fixed scale.  %    
%     OPPURE : at very small distances 
%${\cal D}_{\rm s}=2$.
This follows from the fact that even for a curved metric the spectral sum
(\ref{proofPT}) can be represented by an Euler-Mac Laurin series which always
implies (\ref{proofPT}) as the leading term for $T\rightarrow 0$.

\begin{figure} 
 \begin{center}
 \vspace{0.5cm}
 \hspace{-0.3cm} \includegraphics[height=3.36cm]{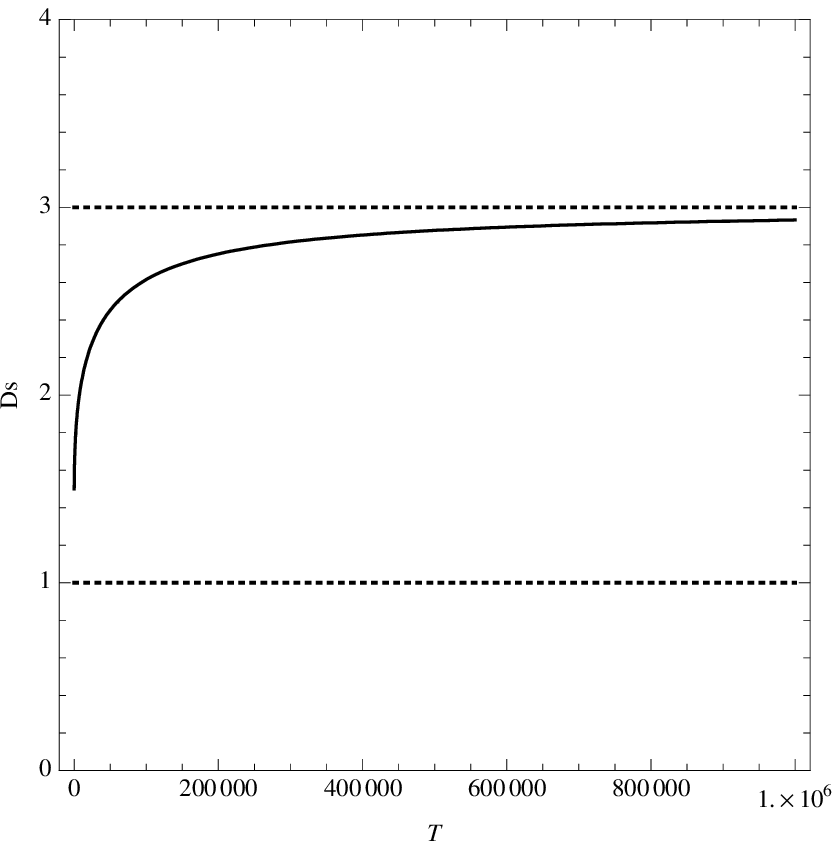}
\includegraphics[height=3.29cm]{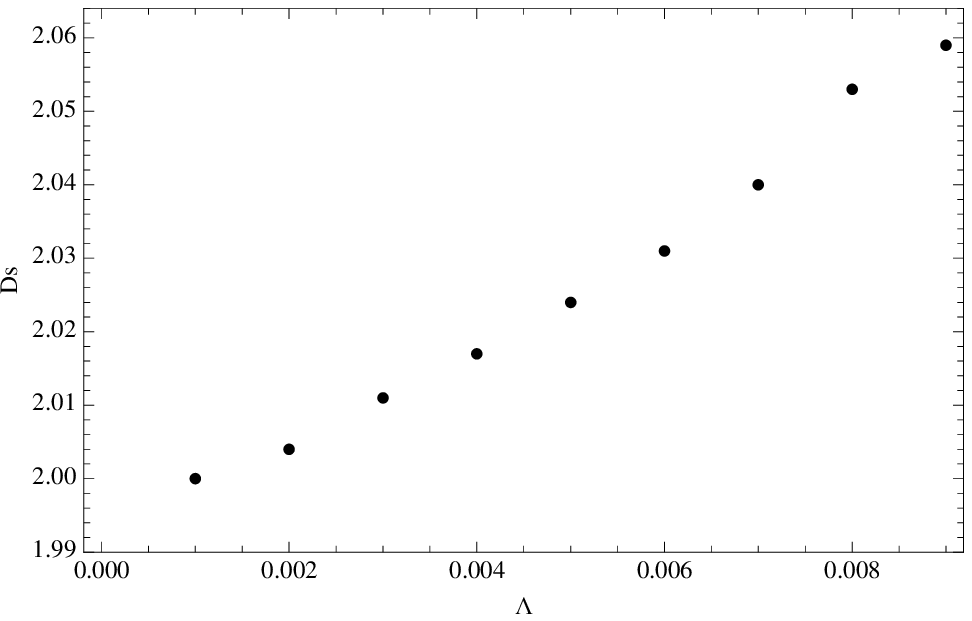}
    \end{center}
    \vspace{-0.0cm}
  \caption{\label{PlotDj6} 
 The plot on the left 
 represents the spectral dimension for $j\gtrsim 4$ or $k\lesssim E_P/4$
 as function of the fictitious time $T$. The dimension at hight energy is $1.5$.
 We have plotted $T \in [0.005, 10^6]$ and used $E_p =1000$, $k_0 =0.01$.
 The plot on the right represents the spectral dimension as
 function of the cosmological constant $\Lambda$ in the Turaev-Viro model
 at the Planck scale ($1 \lesssim j \lesssim 9$).
 }
\end{figure}

Now we have all the ingredients to calculate the spectral dimension using (\ref{trplane})
inside the definition (\ref{specdim}). 
For the PR model
the scaling function ${\mathbb S}(p)$ is obtained 
from (\ref{Sklimits}) replacing $k$ with $p$.
The spectral dimension for $j\gtrsim 4$ or $k\lesssim E_P/4$ increases from 
${\mathcal D}_s \approx 1.5$ to ${\mathcal D}_s \approx 3$ at low energy as it 
is evident from the plot in Fig.\ref{PlotDj6}.
For $1\lesssim j \lesssim 4$ or $E_P \lesssim k \lesssim E_P/4$ using the 
proper scaling we find ${\mathcal D}_s \approx 1.98$. We conclude that the 
fractal dimension decrease from the Planck energy to an intermediate scale where take the 
value $\approx 1.5$ (for $k\approx E_P/4)$ and increase again 
to $\approx 3$ at low energy (Fig.\ref{PlotDj6}).
For the TV model we have differences only for $j \lesssim 4$ and the result
is plotted in Fig.\ref{PlotDj6} on the right.
That plot gives the spectral dimension as a function of the cosmological constant.
The spectral dimension is in the range $2.00 \lesssim {\mathcal D}_s \lesssim 2.059$
for $0.001 \lesssim \Lambda \lesssim 0.009$ in Planck units. In other words the spectral dimension increases with the increase of the cosmological constant at the Planck scale. %in the range $0.004 \lesssim \Lambda \lesssim 0.009$, decreases in the range $0.001 \lesssim \Lambda \lesssim 0.004$ and return to the Ponzano-Regge value for $\Lambda\rightarrow0$.
%\vspace{0.5cm}
%\begin{center}
%fbox{\parbox{15.7cm}{
%\fbox{\parbox{2.6cm}{{\em Proof} of (\ref{trplane})}} 
%}}
%\end{center}
%{\bf End Proof.}% of (\ref{trpropk})}
%\fbox{\parbox{3.5cm}{{\em End proof} of (\ref{trplane})}} 

%Thus we may conclude that on very small and very large length scales the
%spectral dimensions of the QEG spacetimes are
%
%\begin{eqnarray}
%\label{qegspecdims}
%{\cal D}_{\rm s}(T\rightarrow\infty)&=&4\nonumber\\
%{\cal D}_{\rm s}(T\rightarrow 0)&=&2
%\end{eqnarray}
%
%The dimensionality of the fractal realized at sub-Planckian distances is
%found to be 2 again. 

{\em Conclusions and Discussion.}
In this paper we calculated explicitly the spectral dimension (${\mathcal D}_s$)
for the 3d quantum spacetime using the Ponzano Regge spinfoam model.
We considered the simplest decomposition of the spacetime 
and we used the general boundary formalism to characterize the scaling properties 
of the expectation value for the traced propagation kernel.
Using the technical simplifications repeatedly used in the graviton
propagator calculations we have evaluated the nonperturbative 
expectation value of the heat kernel. 

In the PR model and for $k \lesssim E_P/4$ we have plotted ${\mathcal D}_s$ as a function
of a fictitious diffusion time $T$ or equivalently as a function of the length scale.
We obtained three phases: a short scale phase $ l_P \lesssim l \lesssim 4 l_P$ of spectral 
dimension ${\mathcal D}_s \approx 2$, an intermediate scale phase 
$\ell \gtrsim 4 l_P$ of spectral 
dimension ${\mathcal D}_s = 1.5$ and a large scale phase with ${\mathcal D}_s = 3$.

For the TV model the results are equal for $k\lesssim E_P/4$ and to feel the effect 
of the cosmological constant we must goes beyond that energy. The spectral dimension
depends on $\Lambda$ as it is evident from the plot in Fig.\ref{PlotDj6}. %\\\ \\

We interpret the results in the following way. At high energy the spectral dimension
is ${\mathcal D}_s < 3$
because the manifold presents holes typical of an atomic structure.
The cosmological constant basically decreases the number of holes
increasing the spectral dimension. 

{\em Acknowledgements.}
We are extremely grateful to the fantastic environment offered by Perimeter Institute. F. C. 
is in particular indebted with Fotini Markopoulou for inviting him to the Perimeter Institute. 
Research at 
Perimeter Institute is supported by the Government of Canada through Industry Canada 
and by the Province of Ontario through the Ministry of Research \& Innovation.
% Dario Benedetti and James Ryan. %, and Michele Arzano. % 
%for many important and clarifying discussions. 

\end{document}